\documentstyle[12pt]{article}

\newcommand\T{\theta_{12}}
\newcommand\Tb{{\bar\theta}_{12}}
\newcommand\Z{Z_{12}}

\newcommand\D{{\cal D}}
\newcommand\Db{\overline{\cal D}}

\begin{document}
\thispagestyle{empty}
\begin{flushright}
LNF-93/022 (P)\\
BONN-HE-93-18\\
hep-th/9305078 \\
May 1993
\end{flushright}
\begin{center}
{\bf N=2 SUPER BOUSSINESQ HIERARCHY: LAX PAIRS AND CONSERVATION LAWS}
\vspace{1.5cm} \\
S.Bellucci${}^a$, E.Ivanov${}^{b,c}$, S.Krivonos${}^{a,c}$
 and A.Pichugin${}^c$ \vspace{1.5cm}\\
\end{center}
\begin{center}
\noindent ${}^a\;${\it INFN - Laboratori Nazionali di Frascati,
 P.O. Box 13 I-00044 Frascati, Italy} \\
${}^b\;${\it Physikalisches Institut, Universitat Bonn,
Nussallee 12, D-53000 Bonn 1, Germany} \\
${}^c\;${\it Bogoliubov Theoretical Laboratory,
 JINR, Dubna, Head Post Office, P.O. Box 79} \\
$\;\;\;${\it 101000 Moscow, Russian Federation} \vspace{0.5cm} \\
\end{center}
\begin{center}
{\bf Abstract}
\end{center}
We study the integrability properties of the one-parameter family of
$N=2$ super Boussinesq equations obtained earlier by two of us
(E.I. \& S.K., Phys. Lett. B 291 (1992) 63) as a hamiltonian
flow on the $N=2$ super-$W_3$ algebra. We show that it admits
nontrivial higher order conserved quantities and hence gives rise to
integrable hierarchies only for three values of the involved
parameter, $\alpha=-2,\;-1/2,\;5/2$. We find that for
the case $\alpha = -1/2$ there exists a Lax pair formulation
in terms of local $N=2$ pseudo-differential operators, while for
$\alpha = -2$ the associated
equation turns out to be bi-hamiltonian.
\vspace{2cm}
\begin{center}
{\it Submitted to Physics Letters B}
\end{center}
\vfill
\setcounter{page}0
\newpage
\section{Introduction}

During the last few years there has been a considerable interest in
the integrable evolution equations associated with
the $W$-type algebras and superalgebras (see, e.g., [1-8]).
Since the discovery \cite{{Magri},{b2}} that the classical Virasoro
algebra provides
the second hamiltonian structure for the KdV hierarchy, there appeared
a lot of papers where various supersymmetric and $W$ extensions of this algebra
were treated along a similar line and the relevant hierarchies of the
evolution equations were deduced and analyzed. In particular, integrable
$N=1$ \cite{b3}, $N=2$ \cite{a4}, $N=3$ \cite{b4}  and $N=4$
\cite{b5} supersymmetric KdV equations have been constructed, with the
$N=1,2,3,4$ super Virasoro algebras as the second hamiltonian structures.
In \cite{a1} it has been shown that the classical $W_3$ algebra
(with a non-zero central charge) defines a second hamiltonian structure for
the Boussinesq hierarchy. The Lax pair formulation of the latter in terms
of the Gel'fand-Dikii pseudo-differential operators closely related to the
hamiltonian formulation also has been given (see, e.g., \cite{d1}).

Obviously, supersymmetric extensions of the Boussinesq equation
should be associated, in the above sense, with super-$W_3$ algebras. In
ref. \cite{a3} two of us (E.I. \& S.K.) have constructed, in a
manifestly supersymmetric $N=2$ superfield form, the most general
$N=2$ super Boussinesq equation for which the second hamiltonian
structure is given by the classical $N=2$ super-$W_3$ algebra
\cite{Class2W}. This equation turned
out to contain an arbitrary real parameter $\alpha$, much like the
$N=2$ super
KdV equation \cite{a4}.

In this letter we address the question
of existence of the whole $N=2$ Boussinesq hierarchy, i.e. we examine
whether the equation constructed in \cite{a3} admits an infinite sequence
of conserved quantities in involution and a Lax pair formulation.
We find that the nontrivial higher order conserved quantities exist
only for {\it three} values of $\alpha$, namely for $\alpha =
-2,-1/2, 5/2$. This again highly resembles the case of
$N=2$ super KdV equation
which is known to give rise to the integrable hierarchies only for three
special values of the involved parameter \cite{a4}.
We prove the integrability of the option $\alpha = -1/2$ by finding
the Lax pair for it (in terms of the $N=2$ pseudo-differential
operators). We also show that the equation corresponding to the choice
$\alpha=-2$ possesses the first hamiltonian structure. This property,
together with the existence of higher-order conservation laws,
suggest that the $\alpha=-2$ equation is integrable as well.

\section{ N=2 super Boussinesq equation}

Let us first briefly recall the basic points of ref. \cite{a3} in
what concerns the $N=2$ super Boussinesq equation and its relation to
the $N=2$ super-$W_3$ algebra.

All the basic currents of $N=2$ super-$W_{3}$ algebra
\cite{{Class2W},{Romans}}
are accomodated by
the spin 1 supercurrent $J ( Z )$
and the spin 2 supercurrent $T( Z )$, where
$Z \equiv (x, \theta, \bar{\theta})$ are the coordinates of the $N=2,\;1D$
superspace. The supercurrent $J(Z)$ generates the $N=2$ super Virasoro
algebra, while $T(Z)$ can be chosen to be primary with respect to
the latter.
The closed set of SOPE's for these
supercurrents, such that it defines the classical $N=2$ super-$W_3$
algebra, has been written down in \cite{a3}. Here we prefer an equivalent
notation via the  super Poisson brackets
\begin{equation}\label{PB}
\left\{ V_A (Z_1), V_B (Z_2) \right\}_{(2)} =
 \D_{AB}(Z_2) \Delta (Z_{12}) \; ,
\end{equation}
where $V_{A=1,2} \equiv (J,T) \;$ , and $\Delta (Z_{12})$ denotes the
$N=2$ super delta-function
\begin{equation}\label{superdelta}
\Delta (Z_{12}) = \theta_{12}{\bar\theta}_{12}\delta (z_1-z_2) \;.
\end{equation}
The subscript ``(2)'' of the super Poisson brackets
indicates that they provide the second hamiltonian structure for the
$N=2$ super Boussinesq equation to be defined below.

The 2x2 super-differential operator $\D_{AB}$ in (\ref{PB}) encodes
the full information about the structure of the classical
$N=2$ super-$W_3$ algebra.
The explicit form of its entries is as follows:
\begin{eqnarray}
\D_{11} & = & -\frac{c}{8} \left[ \D ,\Db \right]\partial + \Db J \D +
  \D J \Db + J \partial + \partial J \; , \nonumber \\
\D_{12} & = & \Db T \D + \D T \Db + 2 T\partial +\partial T \; ,\nonumber \\
\D_{21} & = & \Db T \D + \D T \Db + 2 T\partial +2\partial T \; ,\nonumber \\
\D_{22} & = & \frac{c}{8}\left[ \D , \Db \right] \partial^3 - 2 J \partial^3 -
   6 \D J \Db \partial^2 - 6 \Db J \D \partial^2 - 6\partial J \partial^2
  - \Db ( 8\partial J -5 T -B^{(2)}) \D\partial     \nonumber \\
 & & -(5T-2\left[ \Db ,\D \right] J + B^{(2)}) \left[ \D ,\Db \right]\partial -
 \D ( 8\partial J +5 T +B^{(2)}) \Db\partial
   \nonumber \\
 & & +\left( \frac{3}{2} \left[ \Db ,\D \right] T -6\partial^2 J +
U^{(3)}\right)\partial
-\frac{1}{2}\partial \left( 5T-2\left[ \Db ,\D \right] J+
  B^{(2)}\right)\left[ \D ,\Db \right]
     \nonumber \\
& & - \left( 3\partial\D T +3\partial^2 \D
J+{\overline{\Psi}}^{(7/2)}\right)\Db
 + \left( 3\partial\Db T -3\partial^2 \Db J-\Psi^{(7/2)}\right)\D \nonumber \\
& &
+\left( -2\partial^3 J+\partial\left[ \Db,\D\right] T
 + \frac{1}{2}\partial U^{(3)} +\frac{1}{2}\Db{\overline\Psi}^{(7/2)}
 + \frac{1}{2}\D \Psi^{(7/2)}-
   \frac{1}{4}\partial \left[ \Db,\D \right] B^{(2)} \right) \; . \label{PB2}
\end{eqnarray}
Here $c$ is the central charge taking an arbitrary value at the classical
level, and
$$B^{(2)}(Z),\; \Psi^{(7/2)}(Z),\;
{\overline\Psi}^{(7/2)}(Z),\;U^{(3)}(Z) $$
are the composite supercurrents with spin
$2,\; {7/2},\;{7/2},\;3 $, respectively,
\begin{eqnarray}
B^{(2)} & = & \frac{8}{c} J^2 \;, \nonumber \\
{\overline\Psi}^{(7/2)} & = & \frac{8}{c} \partial\left( J\D J\right)
   -\frac{72}{c}T\D J +\frac{36}{c}\left[\Db ,\D\right]J \D J +
 \frac{8}{c}J\D T -\frac{128}{c^2}J^2\D J
	    +\frac{4}{c}\partial J\D J \;,\nonumber \\
\Psi^{(7/2)} & = & -\frac{8}{c} \partial\left( J\Db J\right)
   -\frac{72}{c} T\Db J +\frac{36}{c}\left[\Db ,\D\right]J \Db J +
\frac{8}{c}J\Db T -\frac{128}{c^2}J^2\Db J
	    -\frac{4}{c}\partial J\Db J ,\nonumber \\
U^{(3)} & = & \frac{56}{c}J T -\frac{32}{c}J\left[ \Db ,\D \right] J
	 +\frac{128}{c^2}J^3 +\frac{120}{c}\Db J\D J \quad , \label{4}
\end{eqnarray}
where
\begin{equation}\label{5}
\T=\theta_1-\theta_2 \quad , \quad \Tb=\bar\theta_1-\bar\theta_2 \quad ,
 \quad \Z=z_1-z_2+\frac{1}{2}\left( \theta_1\bar\theta_2
-\theta_2\bar\theta_1 \right)
\end{equation}
and the covariant spinor derivatives are defined by
\begin{eqnarray} \label{DefD}
{\cal D}_{\theta} = \frac{\partial}{\partial \theta} -{1\over 2}
\bar{\theta}\partial_x\;,\;\;\;\;&& \bar{{\cal D}}_{\theta} =
\frac{\partial}{\partial \bar{\theta}} -{1\over 2} \theta \partial_x \;,
\nonumber \\
\{ {\cal D}, \bar{{\cal D}} \}=-\partial_x\;,
&&{{\cal D}}^2={\bar {{\cal D}}}^2=0\;.
\end{eqnarray}

The $N=2$ super Boussinesq equation can be defined as the system
of two $N=2$ superfield equations for the supercurrents $T,\;J$ with the
$N=2$ super-$W_3$ algebra (\ref{PB}),
(\ref{PB2}), (\ref{4}) as the second hamiltonian structure.
In other words,
it amounts to the following set of evolution equations:
\begin{equation}\label{b3}
{\dot T}  =  \left\{ T, H \right\}_{(2)} \quad , \quad
{\dot J} =  \left\{ J, H \right\}_{(2)}
\end{equation}
or, in the condensed notation,
\begin{equation}
\dot{V}_A=\D_{AB}\frac{\delta H}{\delta V_B}\;.
\end{equation}
The hamiltonian $H$ in (2.7) and (2.8) is given by
\begin{equation}
H=\int dZ \left( T +\alpha J^2 \right) \; .
 \label{b4}
\end{equation}
We emphasize that (\ref{b4}) is the most general hamiltonian
which can be constructed out of $J$ and $T$ under
the natural assumptions that it respects $N=2$ supersymmetry and has the
same dimension 2 as the hamiltonian of the ordinary bosonic Boussinesq
equation.
Note the presence of the free parameter $\alpha$ in (\ref{b4}).

Now, using the Poisson brackets (\ref{PB}) and the definitions
(\ref{PB2}), (\ref{4}),
it is straightforward to find the explicit form of the $N=2$ super
Boussinesq system:
\begin{eqnarray}
{\dot T} & = & -2 J'''+\left[\Db ,\D\right]  T'+
 \frac{80}{c}\partial\left( \Db J \D J\right)
 -\frac{32}{c}J'\left[\Db , \D\right]J -\frac{16}{c}J\left[\Db , \D\right]J'
+\frac{256}{c^2}J^2J' \nonumber \\
 &+ & \left( \frac{40}{c}-2\alpha \right)\Db J\D  T
     +\left( \frac{40}{c}-2\alpha \right)\D J\Db  T
 +\left(\frac{64}{c}+4\alpha\right)J' T
  +\left(\frac{24}{c}+2\alpha\right)J T' \;, \nonumber \\
{\dot J} & = & 2 T'+\alpha\left( \frac{c}{4}\left[\Db ,\D\right]J'
 +4JJ'\right) \quad .\label{b5}
\end{eqnarray}
The bosonic sector of eqs. (\ref{b5}) is a coupled system of equations
for two spin 2 currents and the spin 1 $U(1)$ Kac-Moody current. As was
shown in ref. [8], the standard Boussinesq equation decouples from
this system only for $\alpha = -4/c$.

Note that the dependence on $c$ in (\ref{b5}) is unessential and
it can be removed by rescaling the superfields $T$ and $J$.
For definiteness, in what follows we will put $c=8$.
On the contrary, the dependence
on $\alpha$ is crucial for achieving integrability: in the next section we
will see that only for three special values of this parameter the above
system results in integrable hierarchies.

\setcounter{equation}0
\section{Conservation laws}

Now we turn to the basic theme of the present paper, the
analysis of integrability of the set (\ref{b5}). The standard signal of
integrability is the presence of an infinite sequence of mutually commuting
nontrivial conserved quantities. In this section we
report on the results of our study of the issue of existence of the higher
order conserved quantities for (\ref{b5}) with $c=8$.

In searching for such objects we made use of the standard method of
undetermined coefficients.
One considers an integral of degree $n$ constructed from all the possible
independent densities of degree $n$, each multiplied by an undetermined
coefficient. (Two densities are dependent if their difference is a total
(super)derivative).
The coefficients are then fixed by requiring the integral to be a
conservation law, that is  time-independent.

In this way, with the heavy use of the symbolic manipulation
program Mathematica \cite{Math}, we
have found the following first six conserved quantities:
\begin{eqnarray}
H_1 & = & \int dZ \; J \;,\nonumber \\
H_2 & = & \int dZ \; \left( T +\alpha J^2 \right)\;, \nonumber \\
H_3 & = & \delta \int dZ \; \left( JT +a_1J^3 +a_2 \D J\Db J \right)
    \;, \nonumber \\
H_4 & = & \int dZ \; \left( T^2 +b_1TJ^2+b_2TJ'+b_3T\D\Db J+
    b_4 J^4 +b_5 J^2\D\Db J +b_6 J J'' \right) \;, \nonumber \\
H_5 & = & \int dZ \; \left( T\D\Db T +c_1 T^2 J +c_2 TJ^3+c_3 TJJ'+
    c_4TJ\D\Db J+c_5T\D J\Db J + c_6 TJ''\right. \nonumber \\
   &   & \left. +  c_7J^5+c_8 J^3\D\Db J +c_9J^2 J'' +c_{10} J^2 \D\Db J'+
      c_{11} J\D\Db J \D\Db J + c_{12} J \D\Db J'' \right)\;, \nonumber \\
H_6 & = & \delta \int dZ \; \left(
     d_1 T^3 +d_2 T T''+d_3 T^2J^2 +d_4 T^2 J'+ d_5 \D T\Db T J+
    d_6 T^2 \D\Db J \right. \nonumber \\
   & & + d_7 TJ^4 + d_8 TJ^2J' +d_9 TJ^2\D\Db J +d_{10}TJ\D J\Db J +
     d_{11} TJJ''+ d_{12}TJ'J' \nonumber \\
  & & + d_{13} TJ'\D\Db J +d_{14} T\D J' \Db J +d_{15} T\D J \Db J'+
      d_{16} T \D\Db J \D\Db J + d_{17} TJ''' \nonumber \\
 & & +d_{18}T\D\Db J''
+ d_{19} J^6 + d_{20} J^4 \D\Db J +d_{21} J^3 J'' +d_{22} J^3 \D\Db J'+
       d_{23} J^2 \D\Db J \D\Db J         \nonumber \\
 & & \left. + d_{24} J^2 J'''
    + d_{25} J^2 \D\Db J''+ d_{26} J\D\Db J \D\Db J'
     +d_{27} JJ^{IV}  \right) \;,
 \label{currents}
\end{eqnarray}
Here
\begin{equation}
\delta \equiv \delta_{\alpha,-2} + \delta_{\alpha,5/2}\;.
\end{equation}

The most striking result of this exercise is that
the nontrivial higher-order  $H_n \; (n \geq 3 )$ exist if and only if the
parameter
$\alpha$ takes one
of the following three values:
\begin{equation}\label{alpha}
\alpha = -2,\; -1/2 ,\; 5/2 \;.
\end{equation}
We have then verified that $H_3$ and $H_6$  exist only for
the two values of $\alpha : \alpha = -2,\;5/2$. Notice that the
special value of $\alpha$ at which the Boussinesq equation in the
bosonic sector decouples from two other equations is present
among those in (\ref{alpha}):
for the choice $c=8$ it is just $\alpha = -1/2$.

The corresponding values of the coefficients in $H_3$ -- $H_6$ are given in
Tables I -- IV.\vspace{0.5cm} \\

\noindent TABLE I. Coefficients of $H_3$. \vspace{0.5cm} \\
\begin{tabular}{|r|r|r|} \hline \hline
$\alpha$ & $a_1$ & $a_2$ \\ \hline
-2 & -5/4 & -5/2 \\
5/2 & 1 & 2 \\ \hline
\end{tabular}
\vspace{0.5cm} \\
TABLE II. Coefficients of $H_4$. \vspace{0.5cm} \\
\begin{tabular}{|r|r|r|r|r|r|r|} \hline \hline
$\alpha$ & $b_1$ & $b_2$ & $b_3$ & $b_4$ & $b_5$ & $b_6$ \\ \hline
    -2 &  -4 & 4   & 8 & 4 & -16 & 4 \\
    -1/2 &  2 & 4   & 8 & -1/2 & -1 & 1 \\
    5/2 &  14 & -8   & -16 & 17/2 & -31 & 7 \\ \hline
\end{tabular}
\vspace{0.5cm} \\
\noindent TABLE III. Coefficients of $H_5$. \vspace{0.5cm} \\
\begin{tabular}{|r|r|r|r|r|r|r|r|r|r|r|r|r|} \hline \hline
$\alpha$ & $c_1$ & $c_2$ & $c_3$ & $c_4$ & $c_5$ & $c_6$ &
$c_7$ & $c_8$ & $c_9$ & $c_{10}$ & $c_{11}$ & $c_{12}$
\\ \hline
 -2 & -3/2  & 15/4 & -10 & -20 & -25/2 & 5/2 & -483/160 & 497/24 & -77/16&
  21/4 & -21/2 & 21/4 \\
-1/2 &  1 & 0   & 0 & 0 & -5 & 0 & 1/5 & 4/3 & -2   & -7/2 & 7 & -1 \\
5/2  & -3/2 & -15/2 & 25/2 & 25 & 10 &
  -5/2 & -33/10 & 113/6 & -13/4 & 21/4 & -21/2 & 4 \\
\hline
\end{tabular}
\vspace{0.5cm}
\noindent TABLE IV. Coefficients of $H_6$. \vspace{0.5cm} \\
\begin{tabular}{|r|r|r|r|r|r|r|r|r|r|} \hline \hline
$\alpha$ & $d_1$ & $d_2$ & $d_3$ & $d_4$ &
$d_5$ & $d_6$ &$d_7$ & $d_8$ & $d_9$  \\ \hline
 -2 & -1/9  & 1/6 &   2/3 & -2/3  &    1 & -4/3   & -4/3  & 20/3   &  40/3 \\
5/2 & 1/117 & 1/26& 23/78 & -7/39 & 5/13 & -14/39 & 31/39 &
-101/39& -202/39 \\ \hline \hline
$d_{10}$& $d_{11}$ & $d_{12}$ & $d_{13} $&
$d_{14}$ & $d_{15}$ & $d_{16}$ & $d_{17}$ & $d_{18}$ & $d_{19}$ \\ \hline
  16 & -10/3 & -8/3 & -16/3 &  -2 &    2  & -16/3 & 2/3   & 4/3  & 8/9 \\
 -55/13 & 1 & 28/39 & 61/39 &5/13 & -5/13 & 61/39 & -2/13 &-4/13 & 5/18 \\
\hline \hline
$d_{20}$ & $d_{21}$ & $d_{22}$ & $d_{23}$ & $d_{24}$ & $d_{25}$ & $d_{26}$ &
$d_{27} $   \\
\cline{1-8}
 -28/3 & 8/3 &-32/9 &   32/3& -16/9 & -14/3 & 20/3 & 2/3     \\
 -89/39 & 115/234 &  -94/117   & 94/39 & -73/234 & -11/13 & 4/3 & 3/26  \\
\cline{1-8}
\end{tabular}
\vspace{0.5cm}\\

The existence of these first higher order nontrivial conservation
laws is a very strong indication of the complete integrability of the
$N=2$ super Boussinesq equation for the three values of
$\alpha$ indicated in eq. (\ref{alpha}) and,
hence, the existence of $N=2$ super Boussinesq hierarchies in these cases.
In the next sections we will present the Lax pair
for the $\alpha=-1/2$ case and the first hamiltonian structure for
the $\alpha=-2$ case.

\setcounter{equation}0
\section{Lax pairs}

In this Section we construct a Lax pair for
the $N=2$ super Boussinesq equation (\ref{b5}) (with $c=8$).

We start from the general multi-parameter form of the third order
Lax operator
\begin{equation}
L  =  \partial^3  + A_1 \cdot \partial^2 +
 A_2 \cdot [\D ,\Db ] \partial + A_3 \cdot  \D\partial +
 A_4 \cdot \Db\partial  + A_5 \cdot \partial + A_6 \cdot [\D , \Db ] +
   A_7 \cdot \D +A_8 \cdot \Db +A_9 \;, \label{lax}
\end{equation}
where $A_1,\ldots, A_{9}$ are arbitrary polynomials in $J,T$ and
their (super)derivatives with suitable dimensions and $U(1)$ properties.
This means, in particular, that $L$ must be a $U(1)$ singlet.
For example, the first several coefficients in $L$ can be parametrized as
follows:
\begin{eqnarray}
A_1  & = & k^{(1)} J \; , \nonumber \\
A_2  & = & k^{(2)} J \; ,\nonumber \\
A_3  & = & k^{(3)} \Db J \; , \nonumber \\
A_4  & = & k^{(4)} \D J \; , \nonumber \\
A_5  & = & k^{(5)}_1 J^2+ k^{(5)}_2 T+
 k^{(5)}_3 \partial J+ k^{(5)}_4 [ \D ,\Db] J \;\;\;\mbox{etc.}\;,
\end{eqnarray}
where $k$ are numerical coefficients.
We search for those values of these parameters for which the Lax equation
\begin{equation}
L_t = \beta \left[ L^{\frac{2}{3}}_{\geq 1} \; , L \right] \label{leq}
\end{equation}
reproduces the $N=2$ super Boussinesq set (2.10).
Here, the subscript $"\geq 1"$ denotes a strictly differential part of
the operator\footnote{We have checked
that the Lax formulation of (\ref{b5}) in a more customary form
$L_t = \beta \left[ L^{\frac{2}{3}}_{+} \; , L \right]$,
where $"+"$ means restriction to the positive and zero parts of the
$N=2$ super pseudo-differential operator $L^{\frac{2}{3}}$, does not
exist at all. The Lax representation of the type (\ref{leq}) was
proposed earlier, e.g., for the $N=2$ super KdV equation in ref. \cite{c1}.}.
For $\beta=-3$
we have found the following two solutions, both corresponding to the same
value of $\alpha = -1/2$:
\begin{eqnarray}
L^{(1)} & = & \partial^3 +3J\partial^2-3\Db J \D \partial+
 \left( 2J^2 -T +\frac{3}{2}\partial J-\frac{1}{2}[\D ,\Db ]J \right)\partial
    \nonumber \\
& & + \left( \Db T -4J\Db J -2 \partial\Db J \right) \D \;,\label{eq1}\\
L^{(2)} & = & \partial^3 +\frac{3}{2}J\partial^2
	   +\frac{3}{2}J [\D ,\Db ]\partial-3\Db J \D \partial+
 \left( J^2 -\frac{1}{2}T +\frac{3}{4}\partial J
   -\frac{1}{4}[\D ,\Db ]J\right)\partial    \nonumber \\
& & + \left( J^2 -\frac{1}{2}T +\frac{3}{4}\partial J
      -\frac{1}{4}[\D ,\Db ]J\right)
  [\D , \Db] + \left( \Db T -4J\Db J -2 \partial\Db J \right) \D\; .
\label{eq2}
\end{eqnarray}
The solutions for the choice $\beta=3$ in (\ref{leq})
can be obtained from (\ref{eq1})-(\ref{eq2}) through the substitutions
$$
J\rightarrow -J \;, \; T\rightarrow T \;,\; \D\rightarrow \Db \;,\;
\Db \rightarrow \D \; .
$$
They yield a Lax pair which is conjugate of (\ref{leq}).

Let us define the $N=2$ super residue of a generic $N=2$ super
pseudo-differential operator
\begin{equation}
{\cal A} = \sum_{ i = -\infty}^M (\beta_i +\gamma_i\D +{\bar\gamma}_i\Db +
     \rho_i [\D ,\Db] )\partial^i
\end{equation}
as the coefficient of $[\D ,\Db]\partial^{-1}$:
\begin{equation}
\mbox{Res } {\cal A} = \rho_{-1} \;.
\end{equation}
Then, following the reasoning of \cite{a4},
we can show that (\ref{leq}) implies the equation
\begin{equation}
\frac{d}{dt} \int \mbox{Res } L^{k/3} d Z =0 \; .
\end{equation}
This gives an infinite number of conservation laws.  We have checked,
that for $L^{(1)}$ all residues of the operators $\mbox{Res }L^{k/3}$
are equal to zero, so $L^{(1)}$ is a
degenerated Lax operator, while for $L^{(2)}$ the expressions
$\mbox{Res }L^{k/3}$ reproduce the conserved quantities for $\alpha =
-1/2$ independently found in the previous section. Note that,
despite the non-self-conjugacy of the operators $L$ and $L^{k/3}$,
the integrals in eq. (4.8) are real: the imaginary parts of the
integrands in all cases prove to be full derivatives.

Thus we have proved the integrability of the $N=2$ super Boussinesq
equation for $\alpha=-1/2$. Its correct Lax form is given by eq. (4.3)
with $\beta = -3$ and the Lax operator $L^{(2)}$ (4.5) (or its conjugate,
with $\beta = 3$ in (4.3)). The Lax operators for the other cases
listed in eq. (\ref{alpha}) (if existing) cannot be represented by
local super-differential operators (nonlocal Lax formulations for
super KdV equations were considered, e.g., in \cite{Pop}).

\setcounter{equation}0
\section{First hamiltonian structure}

In the previous section we have found a Lax pair for the $N=2$ super Boussinesq
equation with $\alpha=-1/2$. Here we study for which values of $\alpha$
the set (\ref{b5}) can be given a first hamiltonian structure.

The first hamiltonian structure for the ordinary Boussinesq
equation can be obtained from the second one by shifting the
stress tensor by a constant. Here we will use the same idea.

In the case at hand there is a substantial freedom compared to the bosonic
case since one can shift the supercurrents
$J$ and $T$ by independent constants.
However, a close inspection of the second Hamiltonian structure (\ref{PB}),
(\ref{PB2}) shows that only shifting the supercurrent $T$ yields a
self-consistent structure:

\begin{equation}\label{PB1a}
\left\{ V_A (Z_1), V_B (Z_2) \right\}_{(1)} = {\widetilde\D}_{AB}(Z_2)
\Delta (Z_{12})\;,
\end{equation}
where now
\begin{eqnarray}\label{PB1b}
{\widetilde\D}_{11} & = & 0 \; , \nonumber \\
{\widetilde\D}_{12} & = & 2\partial \; , \nonumber \\
{\widetilde\D}_{21} & = & 2\partial \; , \nonumber \\
{\widetilde\D}_{22} & = & -5 \left[ \D ,\Db \right] \partial + 7 J \partial +
     9\D J \Db + 9 \Db J \D + 8 \partial J \; . \label{PB1c}
\end{eqnarray}

It is easy to check that this super Poisson structure together with the
proper degree
hamiltonian $H_4$ from the set (\ref{currents}) reproduce the
$N=2$ super Boussinesq
equation only for $\alpha=-2$
\begin{equation}
{\widetilde H}=-\frac{1}{2}\left( H_4\right)_{(\alpha=-2)}=
-\frac{1}{2}\int dZ \left( T^2 -4TJ^2+4TJ'+8T\D\Db J+
 4J^4-16J^2\D\Db J +4 JJ'' \right).
\end{equation}
In this case eq. (\ref{b5}) can be represented in the form which follows
from (2.8) via the substitutions ${\cal D}_{AB} \rightarrow \widetilde
{{\cal D}}_{AB},\; H \rightarrow \widetilde{H}$.
This  proves that our $N=2$ super Boussinesq equation
is bi-Hamiltonian for $\alpha=-2$.
In a number of cases the existence of two hamiltonian structures
for the same equation already implies an infinite
tower of higher order conservation laws (see, e.g., \cite{Magri}).
It would be of interest to see is it true in the present case, i.e.
whether the existence of higher order conserved quantities for
the $\alpha = - 2$ $N=2$ super Boussinesq equation can be traced to
its bi-hamiltonian nature.

The proof of integrability of the $N=2$ super Boussinesq
equation for the remaining value of $\alpha=5/2$ is an open problem.
For this case we were not able to find neither a
Lax pair nor a first hamiltonian structure.

\section{Conclusion}

In this paper we have presented the results of our study of
the integrability properties of the $N=2$ supersymmetric
Boussinesq equation (\ref{b5}) with the $N=2$ super-$W_3$ algebra as
the underlying second hamiltonian structure. We have found that the
integrable $N=2$ super
Boussinesq hierarchies can exist only for the three special
values (\ref{alpha}) of the free parameter $\alpha$. The
integrability in the case $\alpha = -1/2$ stems
from the existence of the Lax pair, while in the case
$\alpha = -2$ it could be a consequence of the presence of
two hamiltonian structures. It is as yet unknown how to account
for the integrability of the case with $\alpha = 5/2$. Perhaps,
one should consider non-local Lax operators
along the lines of ref. \cite{Pop}.

Finally, we wish to point out once more the analogy with the
$N=2$ super KdV equation \cite{a4} which is integrable
also only for the three values of the corresponding free parameter,
$a=-2,4,1$. However, this analogy is not quite literal: e.g., both
cases $a=-2,4$ are known to possess Lax formulations in terms of
local pseudo-differential operators.
The origin of this strange resemblance
is not clear to us, since the underlying second hamiltonian
structures of both systems are essentially different: it is
the $N=2$ super-$W_3$ algebra in the case of super Boussinesq
and the $N=2$ super Virasoro algebra in the case of super KdV.

\vspace{0.5cm}

\noindent{\bf Acknowledgements}

\vspace{0.3cm}

E.I. and S.K. thank Physikalisches Institut in Bonn and LNF-INFN in
Frascati for hospitality during the course of this work. We
are grateful to Z. Popowicz for his critical remarks which helped us
to remove some
doubtful statements in the final version of the paper. This
investigation has been supported in part by the Russian Foundation of
Fundamental Research, grant 93-02-3821.

\vspace{0.5cm}

{\it Note added}. After this paper has been submitted for publication,
we became aware of a preprint by Yung \cite{Yung2} where the existence
of the three distingueshed values of the parameter $\alpha$ has been
also established and the relevant conserved quantities, up to
$H_6$, have been presented.

\end{document}